\documentclass[useAMS,usegraphicx,usenatbib]{mn2e}

\usepackage{graphicx}
\usepackage{color}
\usepackage{epsfig}
\usepackage{float}
\def\aj{AJ}
\def\apj{ApJ}
\def\apjl{ApJ}
\def\aap{A\&A}
\def\mnras{MNRAS}
\def\pasj{PASJ} 
\def\aapr{AAPR} 
\def\nat{Nature}

\newcommand{\teff}{$T_{\rm eff}$}
\newcommand{\kms}{km s$^{-1}$}
\newcommand{\logg}{$\log g$}

\title[]{NGC\,6522: A typical globular cluster in the Galactic bulge without signatures of rapidly rotating Population III stars}

\author[]{Melissa Ness$^1$\thanks{ness@mpia.de},  Martin Asplund$^2$ , Andrew R. Casey$^3$\\
$^1$Max-Planck-Institut f\"ur Astronomie, K\"onigstuhl 17, D-69117 Heidelberg, Germany \\
$^2$ Research School of Astronomy and Astrophysics, Australian National University, Cotter Road, Weston Creek, ACT 2611, Australia \\
$^3$Institute of Astronomy, University of Cambridge, Madingley Road, Cambridge CB3 0HA }
 
\begin{document}

\date{Accepted ? Received ?; in original form April 01, 2014}

\pagerange{\pageref{firstpage}--\pageref{lastpage}} \pubyear{2014}

\maketitle

\label{firstpage}

\begin{abstract}
We present an abundance analysis of eight potential member stars of the old Galactic bulge globular cluster NGC\,6522. 
The same stars have previously been studied by \citet{chiappini2011}, who found very high abundances of the slow neutron capture elements compared with other clusters and field stars of similar metallicity, which they interpreted as reflecting nucleosynthesis in rapidly rotating, massive Population III stars. 
In contrast to their analysis, we do not find any unusual enhancements of the neutron capture elements Sr, Y, Ba and Eu and conclude that previous claims result mainly from not properly accounting  for blending lines. Instead we find NGC\,6522 to be an unremarkable globular cluster with comparable abundance trends to other Galactic globular clusters at the same metallicity ($\mathrm{[Fe/H]}=-1.15\pm0.16$). The stars are also chemically similar to halo and bulge field stars at the same metallicity, spanning  a small range in [Y/Ba] and with normal $\alpha$-element abundances. We thus find no observational evidence for any chemical signatures of rapidly rotating Population III stars in NGC\,6522. 
\end{abstract}

\begin{keywords}
Stars
\end{keywords}

\section{Introduction}

As one of the oldest stellar populations in our Galaxy, globular clusters are key to understanding the earliest epochs of galaxy evolution and stellar nucleosynthesis. Globular clusters may therefore contain unique chemical markers of the first generations of stars. All globular clusters indeed show abundance peculiarities compared with the typical field population, such as an O-Na anti-correlation, which is normally interpreted as reflecting two or more generations of stars in the clusters with the first one sharing the chemical composition with normal field stars of the same metallicity \citep{gratton2012}. 

The nature of the polluter of the later generations of stars is still being debated with intermediate mass asymptotic giant branch stars or rapidly rotating massive stars the leading contenders. As the polluters would be part of the first cluster generation, some of which are still present, they would thus not be metal-free Population III stars but  Population II stars of the same metallicity as the cluster as a whole. Until recently there have been no convincing observations of Population III nucleosynthesis in any globular cluster star but \citet{chiappini2011} claim to have found exactly such a smoking gun in NGC\,6522. 

NGC\,6522 is a little studied globular cluster in the Milky Way's bulge. Observations of individual stars in this cluster are difficult due to the large crowding of more  metal-rich bulge stars and foreground disk contamination in this region as well as the heavy interstellar extinction and reddening characteristic of sight lines of the Galactic bulge, especially near the plane. However, this is a particularly important globular cluster because it may be the oldest globular cluster in the Milky Way \citep{barbuy2009} and it may contain vital clues to the formation and early evolution of the still poorly understood bulge \citep[e.g.][]{ness2013}.  \citet{chiappini2011} recently analysed eight potential NGC\,6522 cluster members and found remarkably high abundances of slow neutron-capture elements, distinct from any other globular cluster observed. The authors demonstrated that such a chemical abundance pattern would be consistent with the nucleosynthesis predicted for rapidly rotating massive, extremely metal-poor or Population III stars, so-called spinstars \citep{pignatari2008}. 

The fact that this cluster is so far unique in carrying such a nucleosynthetic signature may reflect that the very first stars born after the Big Bang formed in the largest over-densities, which subsequently grew with time to become centres of galaxies. The oldest Galactic stars should therefore be concentrated in the bulge region \citep{tumlinson2010}, which also experienced the most rapid chemical enrichment due to the intense star formation rate, making it conceivable that any Population III chemical fingerprint may preferentially show up in an old bulge cluster with $\mathrm{[Fe/H]} \approx -1$ such as NGC\,6522. 

Given the potentially far-reaching implications of the work of \citet{chiappini2011}, we present a re-analysis of the same cluster stars but importantly with a more detailed accounting of blending lines as well as isotopic hyperfine splitting of the $s$-process abundance diagnostics.

\section{Observations}

The GIRAFFE/VLT data from the program 071.B-0617(A) (PI=A. Renzini) were downloaded from the ESO archive. These observations were taken during May-July 2003 and comprised of two field pointings and three wavelength settings: HR13 (612-640nm), HR14 (638.3-662.6nm) and HR15 (679.7-696.5nm). A more comprehensive description of the data can be found in \citet{barbuy2009}.  A number of calibration frames were missing from the ESO archive but were provided by ESO support upon request. We could not locate all of the data corresponding to the total exposure times reported in \citet{barbuy2009} for unknown reasons. The available data were reduced using Gasgano (version 1.24) with the wavelength calibration verified using skylines and telluric lines. 
The data were sky subtracted using an automated python routine that determined  a median sky from the available sky fibres and removed it from each science frame. For each wavelength setting, separate frames were co-added. The resulting signal-to-noise ($S/N$) ratio per resolution element for the three wavelength settings are listed in Table \ref{tab:vels}.

The stellar radial velocities were determined from a cross-correlation with a theoretical model atmosphere spectrum (corresponding to $T_{\rm eff} = 5000$\,K,  $\log g = 3.0$ [cgs], [Fe/H]$=-1.0$) smoothed to the resolving power of the observations using the routine \textsc{fxcor} in \textsc{pyraf}. Bad pixels and telluric regions were masked out for this procedure. 
Our measured radial and heliocentric velocities are listed in Table \ref{tab:vels}.
The mean heliocentric velocity we report for these nine candidate NGC\,6522 stars is $\langle  V_{\rm helio} \rangle = -13.4 \pm 4.9$\,\kms, which is in reasonable agreement with the value quoted by \citet{harris2010}: $-21.1\pm 3.4$\,\kms. It is possible that some of the candidate stars may not be cluster members in spite of the similarity in velocity and metallicity, although there is some discussion in the literature regarding the actual heliocentric velocity of the cluster \citep{terndrup1998}. It should be noted that our heliocentric velocities differ from those reported by \citep{barbuy2009} based on the same observations for unknown reasons. Our inferred heliocentric velocities from the three different wavelength settings obtained at different times are in excellent agreement with each other ($\sigma < 0.7$\,\kms), which is not true for \citet{barbuy2009} (see their Table 3).

\section{Abundance analysis}

After continuum normalisation, we performed a standard 1D LTE abundance analysis using the SMH spectrum analysis program of \citet{casey2013}, which is built around the well-known MOOG code of \citet{sneden1973}. Initial effective temperature guesses were provided by de-reddened 2MASS photometry \citep{strutskie2006} using the \citet{bessell1998} calibration. 
Similarly to \citet{barbuy2009} we discard the star B-134 from further analysis due to contamination of the spectra from a companion/blending star.
Spectroscopic stellar parameters of \teff, \logg\ and [Fe/H] were determined from an iterative procedure to achieve excitation and ionisation balance using $\sim{}$60 Fe\,{\sc i} lines and 12 Fe\,{\sc ii} lines. The microturbulence parameter ($\xi_{\rm turb}$)  was calculated by removing any abundance trends with reduced equivalent width. The final stellar parameters are reported in Table \ref{tab:main}.  The typical uncertainties in the stellar parameters are $\sigma(T_{\rm eff}) = 80$\,K,  $\sigma(\log g) = 0.20$ and $\sigma(\mathrm{[Fe/H]}) = 0.15$, respectively. We report a marginally lower mean metallicity for these eight stars than \citet{barbuy2009}:  $\mathrm{[Fe/H]} = -1.15 \pm 0.16$ compared to $\mathrm{[Fe/H]} = -1.00 \pm 0.20$. 

\

\begin{table*}
 \centering
 \caption{Summary of the measured velocities the nine NGC\,6522 candidate members}
\begin{tabular}{| c  c  c  c  c  c  c  c  c  }
  \hline
Star & Coordinates & Grating & $S/N$  & $V_{\rm rad}$ & $V_{\rm helio}$ & $\langle V_{\rm helio} \rangle$ & $\sigma(V_{\rm helio})$ & [Fe/H]   \\
&  (J2000) &  &  & [\kms]  &  [\kms]  &  [\kms]  &  [\kms] & dex \\
  \hline
\hline
B-008 &18:03:46.04,  -30:00:50.9 & HR13 & 88 & -29.9 & -8.8  & -8.6 & 0.3 & -1.02 \\
 & &  HR14 & 81  & -8.6 & -8.6 & &  & \\
 & &  HR15 & 132 & -16.0  & -8.2  & &  & \\
 \hline
B-107 & 18:03:36.59,  -30:02:16.1 & HR13 & 93 & -27.0 & -5.9 & -6.0 & 0.4 & -1.34\\
 & & HR14 & 86 & -6.5 & -6.5 & &  & \\
 &  & HR15 & 134  & -13.3 & -5.5 & & &\\
 \hline
B-108 & 18:03:35.18, -30:02:04.9 & HR13 & 156 & -33.1 & -12.1  & -13.0 & 0.7 & -1.37 \\
 & & HR14 & 142 &  -13.7 & -13.8 &  & & \\
 & & HR15 & 188 & -20.9 & -13.1 & & & \\
 \hline
B-118 &18:03:42.24,  -30:03:39.9 & HR13 & 87 & -42.9 & -21.9  & -21.9 & 0.1  & -1.04 \\
& & HR14 & 85  & -22.0 & -22.0  & &  &\\
& & HR15& 133 &  -29.7 & -21.9 & &  & \\
 \hline
B-122 & 18:03:33.34, -30:01:58.3 & HR13 & 96  & -36.9 & -15.8  & -15.7 & 0.1 & -0.95 \\
& & HR14 & 92  & -15.6 & -15.7  & &  &\\
& & HR15 & 134 &  -23.3 & -15.5 & &  &  \\
 \hline
B-128 &18:03:44.61, -30:02:10.4 & HR13 & 84 & -33.9  & -12.9  & -12.6 & 0.3& -1.02 \\
& & HR14 & 83  & -12.7 & -12.7  & & & \\
& & HR15& 119 &  -19.9 & -12.1  & & & \\
 \hline
B-130 &18:03:41.00, -30:03:03.0 &HR13& 77 & -36.6 & -15.6  & -15.6 & 0.2 & -1.23\\
& & HR14 & 71 &  -15.3 & -15.4 & & & \\
 & & HR15 & 117 &- 23.7  & -15.9 & &  & \\
\hline
B-134 & 18:03:41.16, -30:02:21.6 & HR13 & 84 & -39.4 & -18.4 &  -18.9 & 0.4 &  --  \\
 & & HR14 & 81 &   -19.3 & -19.4  &  & & \\
 & & HR15 &  127 &  -26.8 & -18.9 & &  & \\
 \hline
F-121 & 18:03:36.41, -30:02:19.8& HR13 & 58 &  -14.1 &  -8.7  & -8.8 & 0.2 & -0.99  \\
 & & HR14 & 93 &   4.3 & -8.6  &  & &  \\
 & & HR15 &  140 &  -7.3 & -9.1 &  & & \\
 \hline
\end{tabular}
\label{tab:vels}
\end{table*}

\begin{table}
 \centering
 \caption{Summary of derived stellar parameters}
\begin{tabular}{| c | c | c | c | c | }
  \hline
Star & \teff  &\logg & [Fe/H] & $\xi_{\rm turb}$ \\
  \hline
\hline
B-008 & 4960 & 2.65 & -1.02 & 2.3 \\
B-107 & 4970 & 1.9 & -1.34 & 2.1 \\
B-108 & 4750 & 2.2 & -1.37 & 1.4 \\
B-118 & 5000 & 2.25 & -1.04 & 2.45 \\
B-122 & 5100 & 2.75 & -0.95 & 1.2 \\
B-128 & 4760 & 2.1 & -1.02 & 1.3 \\
B-130 & 5090 & 2.65 & -1.23 & 2.8 \\
F-121 & 4955 & 2.2 & -1.0 & 1.4 \\
 \hline
\end{tabular}
\label{tab:main}
\end{table}

Individual abundances were determined from the equivalent widths of unblended lines for which reliable oscillator strengths were available from the Gaia-ESO line list (Heiter et al., {\em in prep.}). In the limited wavelength regions of these GIRAFFE observations, we made use of 22 lines of the $\alpha$-elements (Mg, Si, Ca, Ti), two lines available for Al, two for Na, two for Ba, three for Eu, one for La and one for Y.  For the neutron capture elements we performed spectrum synthesis taking care of including isotopic and hyperfine splitting (HFS) and blending lines to ensure the most reliable results. Our mean elemental abundance measurements are given in Table \ref{tab:individual} while the full line-by-line results are provided in Table 4 (online only). We estimate the uncertainties on individual abundances from the quadrature sum of the uncertainties due to the noise in the spectra, and the uncertainties in the stellar parameters and the line-to-line dispersion. These are listed in Table \ref{tab:individual}.

\begin{table*}
 \centering
\caption{Summary of measured abundances for individual elements for the eight candidate NGC 6522 stars. The measurement for each star is in the first row and the upper and lower errors of the measurement are in the two following rows.}
\begin{tabular}{| p {0.8cm} |p {0.8cm} | p {0.8cm} | p {0.8cm}| p {0.8cm} | p {0.8cm} | p {0.8cm} | p {0.8cm} |p {0.8cm} | p {0.8cm} | p {0.8cm} |p {0.8cm} | p {0.8cm} | p {0.8cm} | }
\hline
star & [Fe/H] & [FeI/H] & [FeII/H] & [NaI/Fe] & [MgI/Fe] & [AlI/Fe] & [SiI/Fe] & [CaI/Fe] & [TiI/Fe] & [YII/Fe] & [BaII/Fe] & [LaII/Fe]  & [EuII/Fe]  \\
\hline
B-008  &  --1.02  &  --1.02  &  --1.03 &   0.68  &  0.23 &  1.0  &  0.30  &  0.25  &  0.46   &  0.35  &  0.45  &  0.70   &  0.70  \\
 &   $+$0.15  &  $+$0.15  &  $+$0.15   &  $+$0.1 &  $+$0.2 &  $+$0.1    & + 0.1  &  $+$0.2  &  +0.1   &  $+$0.2 &  $+$0.1 &  +0.15    &  $+$0.15   \\
 &   --0.15  &  --0.15  &  --0.15  &  --0.1 &  --0.2   &  --0.1 &  --0.1  &  --0.2  &  --0.1   &  --0.3 &  --0.1  & -0.2     &  --0.15 \\
B-107  &  --1.34  &  --1.34  &  --1.34  &  0.13 &  0.61 &  0.49  &  0.52  &  0.34  &  0.27     &  0.15  &  0.40  &  0.30  &  0.25  \\
&   $+$0.15  &  $+$0.15  &  +0.15  &  $+$0.25   &  $+$0.2   &  $+$0.25   &  $+$0.15  &  $+$0.15  &  $+$0.1   &  $+$0.25 &  $+$0.1  &   $+$0.2   &  $+$0.15 \\
 &   --0.15  &  --0.15  &  --0.15     &  --0.30 &   --0.25  &   --0.25  &  --0.15  &  --0.15  &  --0.1 &  --0.30 &  --0.1   &  --0.25   &  --0.15  \\
B-108  &  --1.37  &  --1.37  &  --1.37  &  --0.23  &  0.63  &  0.26   &  0.18  &  0.38  &  0.37   &  0.15   &  0.10  &  0.35   &  0.40 \\
 &   $+$0.15  &  $+$0.15  &  $+$0.15  &  $+$0.15  &  $+$0.15 &  $+$0.15  &  $+$0.15  &  $+$0.15  &  $+$0.15   & --0.05  &  $+$0.15  &  $+$0.25    &  $+$0.15  \\
  &   --0.15  &  --0.15  &  --0.15   &  --0.15  &  --0.15  &  --0.40  &  --0.40  &  --0.15  &  --0.15 &  -0.05  &  --0.15  &  -0.45   &  --0.2  \\
B-118  &  -1.04  &  -1.04  &  -1.04  &  0.40  &  0.37  &  1.01 &  0.29  &  0.28  &  0.54   &  0.30 &  0.30  &  0.55     &  0.40   \\
  &   $+$0.15  &  $+$0.15  &  $+$0.15   &  $+$0.1  &  $+$0.1 &  $+$0.1  &  $+$0.1  &  $+$0.15  &  $+$0.15    &  $+$0.2 &  $+$0.1  &  $+$0.1   &  $+$0.15  \\
 &   --0.15  &  --0.15  &  --0.15  &  --0.1  &  --0.1  &  --0.1 &  --0.1  &  --0.15  &  --0.15  &  --0.30  &  --0.1  &  --0.1     & --0.15  \\
B-122  &  --0.95  &  --0.95  &  --0.95  &   0.42  &  0.36  &  0.62 &  0.21  &  0.29  &  0.33   &  0.10  &  0.35  &  0.30    &  0.50 \\
 &   $+$0.15  &  $+$0.15  &  $+$0.15 &  +0.1  &  $+$0.15 &  $+$0.1  &  $+$0.15  &  $+$0.15  &  $+$0.1  &  $+$0.25  &  $+$0.1  &  +0.1     &  $+$0.15 \\
 &   --0.15  &  --0.15  &  --0.15    &  --0.1  &  --0.15  &  --0.1 &  --0.15  &  --0.15  &  --0.1  &  --0.35 &  --0.1  &  --0.15   &  --0.15  \\
B-128  &  --1.02  &  --1.03  &  --1.03  &  0.20 &  0.34  &  0.82  & 0.21  &  0.50  &  0.25   &  0.40  &  0.80  &  0.30  & 0.0 \\
 &   $+$0.1  &  $+$0.1  &  +0.1  &  $+$0.15  &  +0.15 &  $+$0.1  &  $+$0.15  &  $+$0.15  &  $+$0.1  &  $+$0.2   &  $+$0.1  &  $+$0.15   &  $+$0.15  \\
 &   --0.1  &  --0.1  &  --0.1    &  --0.15   &  --0.15 &  --0.1 &  --0.15  &  --0.15  &  --0.1  &  --0.25  &  --0.1   &  --0.15    &  --0.15 \\
B-130  &  --1.23  &  --1.24  &  --1.24  &  0.46  &  0.34  &  0.64  &  0.32  &  0.35  &  0.40  &  0.44 &  0.10  &  0.45      &  0.90 \\
 &   $+$0.15  &  $+$0.15  &  $+$0.15   &  $+$0.15  &  $+$0.25  &  $+$0.2  &   $+$0.2  &  $+$0.15  &  $+$0.1 &  $+$0.25   &  $+$0.1  &  $+$0.2   &  $+$0.15  \\
  &   --0.15  &  --0.15  &  --0.15    &  --0.25  &  --0.25   &  --0.2 &  --0.2  &  --0.15  &  --0.1  &  --0.30  &  --0.1  &  --0.2   &  --0.15 \\
F-121  &  --1.0  &  --0.99  &  --1.01   &  -0.11  &  0.54  &  0.12  &  0.20  &  0.43  &  0.42   &  0.20 &  0.45  &  0.10   &  0.30  \\
 &   $+$0.15  &  $+$0.15  &  $+$0.15   &   $+$0.20  &  $+$0.2  &  $+$0.2 &  $+$0.15  &  $+$0.15  &  $+$0.15   &  $+$0.2  &  +0.2  &  $+$0.1    &  $+$0.10 \\
  &   --0.15  & --0.15  &  --0.15    &  --0.3  &  --0.15 &  --0.2  &  --0.15  &  --0.15  &  --0.15    &  --0.25 &  --0.2  &  --0.1    &  --0.15 \\
  \hline
\end{tabular}
\label{tab:individual}
\end{table*}

In general we find comparable stellar abundances to \citet{barbuy2009}, with marginally lower [Fe/H] (by $0.15$\,dex) and slightly higher [$\alpha$/Fe] ratios (by $\sim 0.2$\,dex).
These differences can likely be traced to the adopted line-lists and difference in adopted stellar parameters (our $T_{\rm eff}$ and $\xi_{\rm turb}$ are typically 200\,K and 0.6\,\kms, respectively higher). We also report roughly similar abundances for [Na/Fe], [La/Fe] and [Eu/Fe]  (which have mean values of $+0.2$ dex, $+0.4$ dex and $+0.4$ dex)  but lower Ba abundance compared to \citet{barbuy2009}. The inclusion of hyperfine splitting makes only a small difference to the measurements. Our mean measurement for Ba is $\mathrm{[Ba/Fe]} = 0.35 \pm 0.2$, compared to $\mathrm{[Ba/Fe]} = 0.5 \pm 0.4$ in the original analysis. 

\begin{figure}
\centering
\includegraphics[scale=0.18]{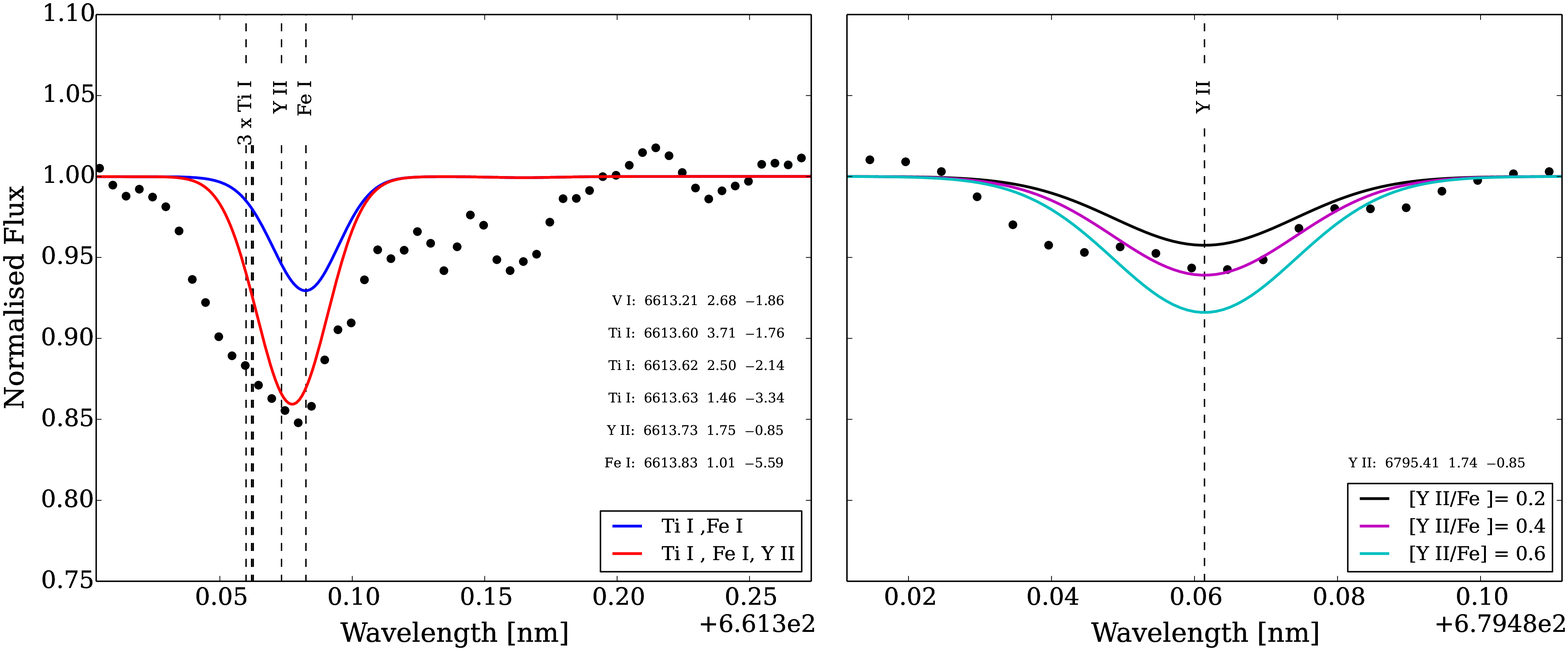}
\caption{The blended Y II feature, which is very similar for all stars, is shown for star B-128. The panel at left shows the synthesis of this line with the contribution of Ti I and Fe I in blue and the contribution of Ti I, Fe I and Y II in red, for the Y value measured for this star shown at right, of [Y/Fe] = 0.4 dex (and [Ti/Fe] = 0.25 as measured from the Ti lines in the spectra). The panel at right shows the best fit abundance of  [Y/Fe] = 0.4 dex and syntheses at $\pm$ 0.2 dex comparison, for the unblended feature at 679.4nm which we use to derive all of our [Y/Fe] abundances. The parameters of the absorption features are provided in the figures, to indicate the relative line strengths, in order of element, wavelength, excitation potential and log $gf$} 
\label{fig:Y}
\end{figure}

A more striking discrepancy is obtained for Y compared to the analysis of \citet{chiappini2011}. Our mean ratio is $\mathrm{[Y/Fe]} = 0.25 \pm 0.10$ compared to $\mathrm{[Y/Fe]} = 1.2 \pm 0.2$ in their analysis. Here the difference is largely due to adopted Y lines. In our analysis, we use a single unblended Y\,{\sc ii} line at  679.54\,nm, whereas the original analysis relied on the Y\,{\sc ii} line at 661.37\,nm, which they did not realise is severely blended by in particular a  Fe line at 661.38nm and a Ti line at 661.36 nm  (see Figure \ref{fig:Y}). We obtain a higher abundance of [Y/Fe] by about +0.6 dex when fitting for [Y/Fe] at 661.37\,nm without accounting for the blending lines.

\begin{figure}
\centering
\includegraphics[scale=0.3]{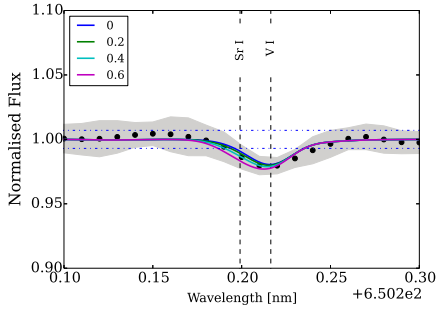}
\caption{The median spectra from all 8 stars combined for the Sr feature at 650.4nm. The synthesis shows the line profiles for abundance ratios of [Sr/Fe]  = 0,0.2,0.4,0.6 for the typical [V/Fe] ratio of [V/Fe] $\approx$ 0 for the stars. The solid grey lines represent the 1$-\sigma$ measurement of the mean of the 8 stars and the dashed lines either side of the line profile show the width of the noise of the combined spectra. The best fit to the line adopting [V/Fe] = 0 is [Sr/Fe] = 0.40. } 
\label{fig:Sr}
\end{figure}

Finally, we argue that no reliable Sr abundance can be inferred from these individual GIRAFFE spectra. \citet{chiappini2011} relied on the Sr\,{\sc i} 650.40\,nm line, which in our opinion is not discernible over the noise of the individual spectra and is blended. This feature, for the median combination of all 8 stars is shown in Figure \ref{fig:Sr}, synthesised with 0 -- 0.8 dex of [Sr/Fe] enhancement and a fixed [V/Fe] = 0. A good fit is obtained with an [Sr/Fe] = 0.4 given a [V/Fe] = 0.  The four Sr lines in the Gaia-ESO line-list available in this wavelength region (640.85\,nm, 655.02\,nm, 654.68\,nm, 661.73\,nm, 671.91\,nm) are also too weak for abundance purposes. As a result we do not quote any [Sr/Fe] values.

\begin{figure}
\centering
\includegraphics[scale=0.35]{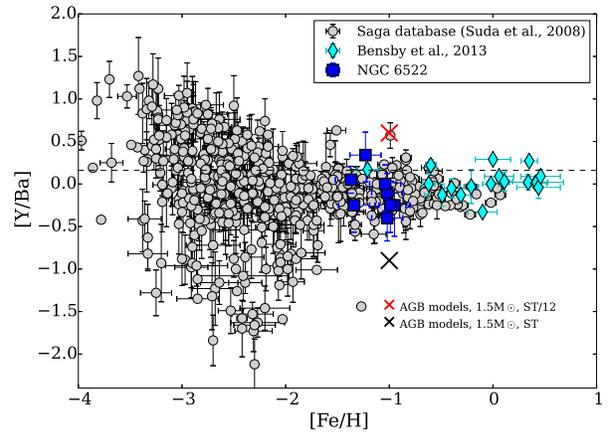}
\caption{[Y/Ba] ratios for stars of the Galactic halo from the SAGA database (grey circles) \citep{suda2008} and bulge stars of Bensby et al., (2013) (cyan diamonds) compared to the eight candidate NGC\,6522 stars (blue squares). Typical model predictions for the contribution of the main r-process is shown at 0.16 dex, along the dashed black line and the model predictions for a standard (ST) and reduced efficiency ST/12 (factor 12) AGB models are shown in the red and black crosses, respectively (see Figure 1 from \citet{chiappini2011}.}
\label{fig:yba}
\end{figure}

\begin{figure*}
\centering
\includegraphics[scale=0.38]{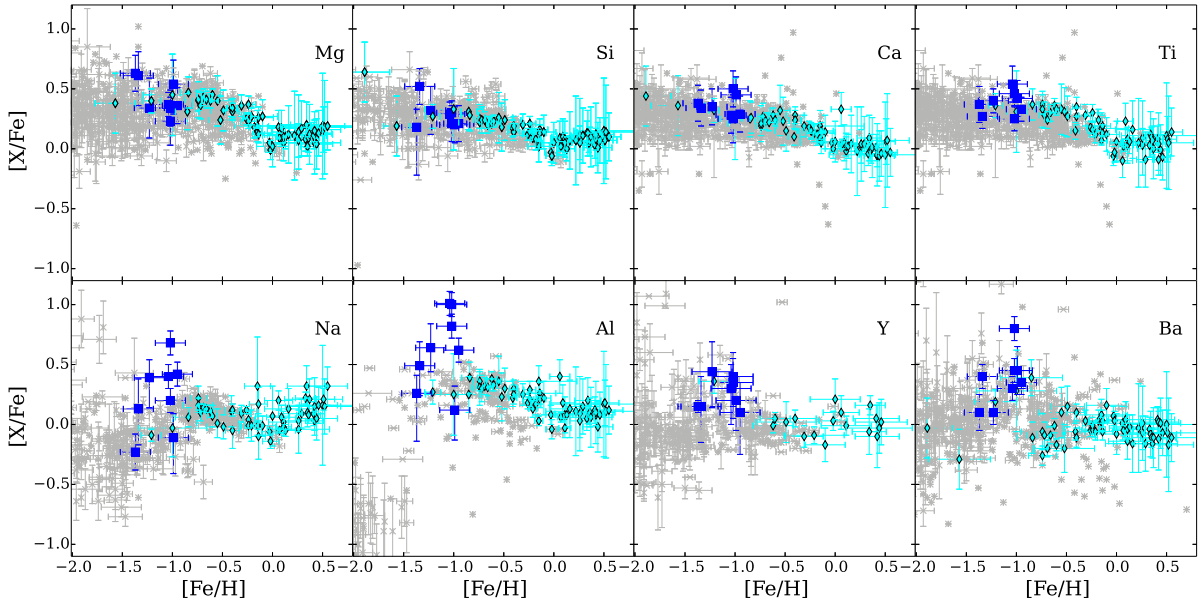}
\caption{Comparison of Na, Mg, Ca, Si, Ti, Al, Ba and Y abundances measured for the eight stars (blue squares) compared with stars of the bulge \citep{bensby2013} (cyan triangles) and halo stars of the SAGA database (grey crosses).}
\label{fig:bulge}
\end{figure*}

Figure \ref{fig:bulge} compares our inferred abundance ratios for NGC\,6522 to micro-lensed dwarf stars of the bulge from \citet{bensby2013} as well as typical halo stars \citep{suda2008}, demonstrating in general a good agreement.  The cluster stars show comparable abundances to the bulge for Mg, Si, Ca, Ti, Y and Ba, while the scatter is Na and Al is directly attributable to the Na-Al correlation typical for globular clusters. Our [Y/Fe] and [Ba/Fe] ratios are consistent with other globular clusters at this metallicity \citep[e.g.][]{johnson2010}. 
Figure \ref{fig:yba} shows the [Y/Ba] of the eight NGC\,6522 stars compared to field bulge and halo stars. The cluster stars are not anomalous compared to field stars of similar metallicity, neither in [Y/Ba] nor in their scatter, in contrast to the conclusion by \cite{chiappini2011}.

\section{Conclusions}

We have re-analysed available GIRAFFE/VLT spectra of eight members of the bulge globular cluster NGC\,6522 from the ESO archive. We have paid particular attention to considering the effects of isotopic hyperfine splitting and blending lines on the derived abundances of the $s$-process elements. We also use in some cases on more reliable, cleaner lines. As a result our results are systematically and significantly lower than those reported by \citet{chiappini2011}, which did not consider those effects.

In all cases the NGC\,6522 stars display elemental abundances typical of field halo and bulge as well as other globular clusters of similar metallicities. 
We conclude that neither the normal $s$-process abundances nor the low and fairly constant [Y/Ba] ratios measured in NGC\,6522 support a nucleosynthetic origin by rapidly rotating Population III stars. In terms of its chemical composition, NGC\,6522 is an unremarkable globular cluster.

\section*{Acknowledgments}
We appreciate helpful and stimulating discussions with Beatriz Barbuy and Cristina Chiappini. 
The research has
received funding from the European Research Council under the European
Union's Seventh Framework Programme (FP 7) ERC Grant Agreement n.
[321035]. MA acknowledges generous funding from the Australian Research Council (grant FL110100012)
MA acknowledges generous funding from the Australian Research Council (grant FL110100012). ARC acknowledges funding from European Research Council grant 320360: The Gaia-ESO Milky Way Survey.


\end{document}